\documentclass{PoS}
\usepackage{epsfig}
\usepackage{latexsym}
\usepackage{amsmath}
\usepackage{amsfonts}
\newcommand{\Tr}{\mathop{\textrm{Tr}}}
\renewcommand{\Re}{\mathop{\textrm{Re}}}
\newcommand{\Det}{\mathop{\textrm{Det}}}
\newcommand{\Nc}{N_{\rm c}}

\PoS{PoS(LAT2005)189}

\title{Four-loop plaquette in 3d with a mass regulator}

\ShortTitle{Four-loop plaquette in 3d with a mass regulator}

\author{\speaker{Christian Torrero}\\
        University of Parma and INFN\\
        E-mail: \email{torrero@fis.unipr.it}}

\author{Francesco Di Renzo\\
        University of Parma and INFN\\
        E-mail: \email{direnzo@fis.unipr.it}}
        
\author{Vincenzo Miccio\\
        University of Milano Bicocca and INFN\\
        E-mail: \email{vincenzo.miccio@mib.infn.it}}      

\author{Mikko Laine\\
        University of Bielefeld\\
        E-mail: \email{laine@physik.uni-bielefeld.de}}        
        
\author{York Schr\"oder\\
        University of Bielefeld\\
        E-mail: \email{yorks@physik.uni-bielefeld.de}}

\abstract{The QCD free energy can be studied by dimensional
reduction to a three-dimensional (3d) effective theory, whereby
non-perturbative lattice simulations become less demanding. To connect 
to the original QCD a perturbative matching computation is required, 
which is conventionally carried out in dimensional regularization. 
Therefore the 3d lattice results need to be converted to this 
regularization scheme as well. The conversion must be carried 
up to 4-loop order, where the free energy displays an
infrared (IR) singularity. We therefore need a regulator which
can be implemented both on the lattice and in the continuum 
computation. We introduce a mass regulator to perform Numerical
Stochastic Perturbation Theory computations. Covariant gauge is fixed
in the Faddeev-Popov scheme without introducing any ghost fields.}

\FullConference{XXIIIrd International Symposium on Lattice Field Theory\\
                25-30 July 2005\\
                Trinity College, Dublin, Ireland}

\begin{document}

%%%%%%%%%%%%%%%%%%%%%%%%%%%%%%%%%%%%%%%%%%%%%%%%%%%%%%%%%%%%%%%%%%%%%%%%%%%%%
%
\section{Introduction}

As is well known, finite-temperature QCD seems to show two 
different phases: it is confining at low temperatures (the realm of
mesons and baryons) while asymptotic freedom and a quark-gluon plasma 
are expected to appear in the high-temperature regime.
A good observable to witness the change
is the QCD free energy density, given essentially
by the familiar Stefan-Boltzmann law of blackbody radiation, 
multiplied by the number of light effective degrees of freedom. 

To study the free energy density requires different methods in 
different regimes. At low temperatures the problem
has to be treated with numerical lattice simulations, while at high
temperatures perturbation theory should allow at least for some 
progress, given that the coupling constant $g$ is small.
Nevertheless, even for small $g$, certain coefficients
in the weak-coupling expansion do remain non-perturbative~\cite{linde}, 
and can only be determined with numerical techniques. 

In the high-temperature regime, the theory contains
three different momentum scales~\cite{dr},
namely $T$ (hard modes), $gT$ (soft modes) and $g^2T$ (ultrasoft modes). 
The contribution of each of these modes
is best isolated in an effective theory setup. This is accomplished
via \emph{dimensional reduction}~\cite{dr,generic,bn} 
by integrating out the hard and soft modes to 
obtain a 3d pure Yang-Mills SU(3) theory (``MQCD''). 
MQCD can then be analysed on the lattice and the results can 
be added to the various perturbative contributions
to obtain the complete answer. 

To add the MQCD lattice results to the perturbative ones, we
need to change regularization scheme from lattice to dimensional
regularization. To this aim, a matching between lattice and
continuum computations is needed and this is achieved by means of
Lattice Perturbation Theory applied to MQCD.
The strategy we adopt for this purpose here is the one of
\emph{Numerical Stochastic Perturbation Theory} (NSPT) 
developed in recent years by the Parma group.

%%%%%%%%%%%%%%%%%%%%%%%%%%%%%%%%%%%%%%%%%%%%%%%%%%%%%%%%%%%%%%%%%%%%%%%%%%%%%
%
\section{The NSPT method}

NSPT relies on \emph{Stochastic Quantization}~\cite{pw} which 
is characterized by the introduction of an extra coordinate, 
a stochastic time $t$, together with an evolution equation called 
the Langevin equation, 
\begin{equation}
 \frac{\partial\phi(x,t)}{\partial t}=
  -\frac{\partial S[\phi]}{\partial\phi}+\eta(x,t)~,
\end{equation}       
where $\eta(x,t)$ is a Gaussian noise which effectively generates 
the quantum fluctuations of the theory. 

The average over this noise is such
that, together with the appropriate limit in $t$, 
the desired Feynman-Gibbs functional integration is reproduced: 
\begin{equation}
 \big\langle O[\phi_{\eta}\small(x,t\small)]\big\rangle_{\eta}
 \stackrel{t\rightarrow\infty}{\bold\longrightarrow} 
 \frac{1}{Z}\int [D\phi] 
  O[\phi\small(x\small)]e^{-S[\phi(x)]} \;.
\end{equation}   
For SU(3) Yang-Mills theory, the Langevin equation becomes
\begin{equation}
 \partial_{t}U_{\eta} = -i\Bigl( \nabla S[U_{\eta}]+\eta \Bigr) U_{\eta} 
 \;,
\end{equation}
guaranteeing the proper evolution of variables within the group. 

In this framework, perturbation theory comes into play by means
of the expansion~\cite{fdr1}
\begin{equation}
 U_{\eta}(x,t)\longrightarrow\sum_{k}g_0^kU_{\eta}^{(k)}(x,t) 
 \;,
\end{equation} 
where $g_0$ is the bare gauge coupling. 
This results in a system of coupled equations that can be numerically
solved via a discretization of the stochastic time $t=n\tau$, 
where $\tau$ is a time step. In
practice, we let the system evolve according to the Langevin equation
for different values of $\tau$, average over each
thermalized signal (this is the meaning of the above-mentioned limit
$t\rightarrow \infty$), and then extrapolate in order to get the
$ \tau = 0 $ value of the desired observable. This procedure is then
repeated for different values of the various parameters appearing in
the action.\\

%%%%%%%%%%%%%%%%%%%%%%%%%%%%%%%%%%%%%%%%%%%%%%%%%%%%%%%%%%%%%%%%%%%%%%%%%%%
%
\section{Mass as an IR regulator}
 
As stated above, the quantity we are interested in is the
contribution to the QCD free energy density $f$ coming from the 3d pure
SU(3) theory. On the lattice, this observable is related to
the trace of the plaquette $\langle 1-\Pi_{P}\rangle$,
where $\Pi_P \equiv \Nc^{-1}\Re\Tr P$ and $P$ is the elementary plaquette, via 
\begin{equation}
 \langle 1-\Pi_{p} \rangle
 =\frac{2a^d}{d(d-1)}\frac{\partial}{\partial \beta_0}
 \Bigg(\frac{f}{T} \Bigg) 
 \;,
\end{equation}   
with bare lattice coupling $\beta_0=2N_c/(a^{4-d}g_0^2)$. 
The outcome can be expanded in powers of $\beta_0$ as 
\begin{equation}
 \langle 1-\Pi_{p}\rangle=
 \frac{\displaystyle c_{1}}{\displaystyle\beta_{0}}+
 \frac{\displaystyle c_{2}}{\displaystyle\beta_{0}^2}+
 \frac{\displaystyle c_{3}}{\displaystyle\beta_{0}^3}+
 \frac{\displaystyle \widetilde{c}_{4}}{\displaystyle\beta_{0}^4}+
 {O}(\displaystyle\beta_{0}^{-5})~.
\end{equation}
The determinations of the 
first three coefficients in the present setting
have been discussed in Ref.~\cite{fdr2}.
The non-perturbative value of the whole quantity has
been determined with lattice simulations in Ref.~\cite{plaq}.
Terms of ${O}(\displaystyle\beta_{0}^{-5})$ disappear
in the continuum limit, thanks to the super-renormalizability
of the theory. Thus only the fourth order coefficient is
missing at the moment. 

As shown parametrically in Ref.~\cite{linde} and 
explicitly in Refs.~\cite{gsixg}, the coefficient 
$\widetilde{c}_4$ is actually IR divergent, and
consequently an appropriate regulator must be introduced 
for its determination. In a non-perturbative setting this 
is provided by confinement, while in fixed-order 
computations one could employ a finite volume (as in Ref.~\cite{fdr2})~ or a mass. 
Since the use of a mass is more convenient in continuum 
computations involving dimensional regularization, 
we need to implement it in lattice perturbation theory as well. 

Apart from introducing a mass, we also fix the gauge in order 
to match the setting of the continuum computations. 
Consequently, the functional integral is given by
\begin{equation}
 Z=\int\! [D\phi] \,
 {\Det\Bigl(-\sum_{\mu} \hat{\partial}_{\mu}^{L}\hat{D}_{\mu}[\phi]+m^2\Bigr)}
 \exp\big(-S_{\scriptscriptstyle{W}}-S_{\scriptscriptstyle{GF}}\big)=
 \int\! [D\phi]\,
 \exp\big(-S_{\scriptscriptstyle{W}}-S_{\scriptscriptstyle{GF}}-
  S_{\scriptscriptstyle{FP}}\big)\;,
\end{equation}
where we assume the use of lattice units (i.e. $a=1$), and 
\begin{eqnarray}
 S_{\scriptscriptstyle{W}} 
 & = & \beta_{0}\sum_{P}(1-\Pi_{P})
 +\frac{\beta_{0}m^2}{4\Nc}\sum_{x,\mu,A}\phi_{\mu}^{A}(x)\phi_{\mu}^{A}(x)
 \;, \\
%\end{equation}
%\begin{equation}
 S_{\scriptscriptstyle{GF}} 
 & = & \frac{\beta_{0}}{4\Nc\alpha}\sum_{x,A}\Bigl[\sum_{\mu}
 \hat{\partial}_{\mu}^{L}\phi_{\mu}^{A}(x)\Bigr]^2 
 \;, \\
%\end{equation}
%\begin{equation}
 S_{\scriptscriptstyle{FP}} 
 & = & -\Tr\Bigl[\ln\Bigl(-
 \sum_{\mu}\hat{\partial}_{\mu}^{L}\hat{D}_{\mu}[\phi]+m^2\Bigr)\Bigr]
 \;,
\end{eqnarray}   
where we have followed the conventions of Ref.~\cite{hjr}, 
writing in particular 
$U_\mu = \exp(i \phi_\mu)$, 
$\phi_\mu = \phi_\mu^A T^A$,
with the normalization $\Tr[T^AT^B] = \delta^{AB}/2$.
Moreover $m$ is the common gluon and ghost mass, $\alpha$ is the gauge
parameter, and $\hat D_{\mu}$ is the discrete Faddeev-Popov operator, 
given by~\cite{hjr}
\begin{equation}
 \hat{D}_{\mu}[\phi]=
 \biggl[1+\frac{i}{2}\displaystyle \Phi_{\mu}
 -\frac{1}{12} \displaystyle \Phi_{\mu}^2
 -\frac{1}{720} \displaystyle \Phi_{\mu}^4
 -\frac{1}{30240}\displaystyle \Phi_{\mu}^6
 +{O}(\displaystyle \Phi_{\mu}^8)\biggr]
  \hat{\partial}_{\mu}^{R}+i\displaystyle \Phi_{\mu}
 \;,
\end{equation}
with $\displaystyle \Phi_\mu = \phi_\mu^A F^A$, where $[F^A]_{BC} \equiv - i f^{ABC}$
are the generators of the adjoint representation. 

To treat the Faddeev-Popov determinant as a part of the action means
that, because of the Langevin equation, one has to face the quantity
~$\nabla S_{\scriptscriptstyle{FP}}=-\nabla \Tr[\ln B]=
-\Tr[\nabla BB^{-1}]$ with
$B[\phi]=-\sum_{\mu}\hat{\partial}_{\mu}^{L}\hat{D}_{\mu}[\phi]+m^2$.
We perform the 
inversion as in Ref.~\cite{fdr3}, while the trace is
computed by means of sources in the usual way. 

The global strategy is then to
perform simulations with lattices of different sizes at fixed mass in
order to extrapolate to infinite volume and, afterwards, to repeat
this procedure for other values of the mass. At this point, after
subtracting the expected logarithmic divergence, one extrapolates to
zero mass, obtaining the needed fourth order coefficient. It is crucial
to take the infinite-volume limit before the zero-mass one because, by
performing the limits in the opposite order, the final IR regulator
would be the volume and not the mass as we want.

%%%%%%%%%%%%%%%%%%%%%%%%%%%%%%%%%%%%%%%%%%%%%%%%%%%%%%%%%%%%%%%%%%%%%%%%%
%
\section{First (benchmark) results}

So far, the statistics we took are not sufficient to carry out
the infinite-volume and zero-mass limits for the
fourth order coefficient $\tilde c_4$, but it is already 
possible to crosscheck the reliability of the general method. 
As a first test, we
compare the 1-loop numerical results for the trace of the
plaquette for the various masses with the known analytic values. 
As shown in Fig.~1 for a lattice extent $L= 5$, the agreement between
the numerical values and the analytic curve is very good. 

A second check could consist of extrapolating at fixed lattice extent
to zero mass, to see if one recovers the already
known coefficients~\cite{fdr2}. Figs. 2 -- 5 show these extrapolations 
for a lattice extent $L= 7$: the fitting curve is a polynomial in $m^2$
(the most naive choice) and it seems to approach the expected result
(the point at $m=0$) very well for all the loop orders. The
numerical values are given in Table 1.
Both of the mentioned checks are well satisfied also 
for the other lattice extents that we have employed so far.
%
%%%%%%%%%%%%%%%%%%%%%%%%%%%%%%%%%%%%%%%%%%%%%%%%%%%%%%%%%%%%%%%%%%%%%%%%
%
\begin{table}[h]
\begin{center}
\begin{tabular}{|r|r|r|}
\hline
Loop & Result from a fit to $m=0$ & Direct measurement at $m=0$ \\
\hline
1 & -2.6594(17) &  -2.6580(8) \\
\hline
2 & -1.9166(63) &  -1.9095(30) \\
\hline
3 & -6.304(37)  &  -6.307(21) \\
\hline
4 & -28.43(27)  & -28.68(15) \\
\hline
\end{tabular}
\caption{Comparison of the zero-mass extrapolations 
with the known results~\cite{fdr2} (lattice extent $=7$).}
\label{Tab.1}
\end{center}
\end{table}
%%%%%%%%%%%%%%%%%%%%%%%%%%%%%%%%%%%%%%%%%%%%%%%%%%%%%%%%%%%%%%%%%%%%%%%%%%%%%

As for the 4-loop order, Fig. 6 shows the behavior  
with respect to the lattice size at fixed mass: the result seems 
to stabilise towards the infinite-volume value in the way one would 
expect. Once a few more lattice sizes are available and similar
extrapolations can be carried out for all masses, we will
finally be in a position to carry out the mass extrapolation
that is our ultimate goal. 

%%%%%%%%%%%%%%%%%%%%%%%%%%%%%%%%%%%%%%%%%%%%%%%%%%%%%%%%%%%%%%%%%%%%%%%%%%%%
%
\begin{figure}[h]
 \hfill
 \begin{minipage}[t]{.45\textwidth}
  \begin{center} 
   \epsfig{file=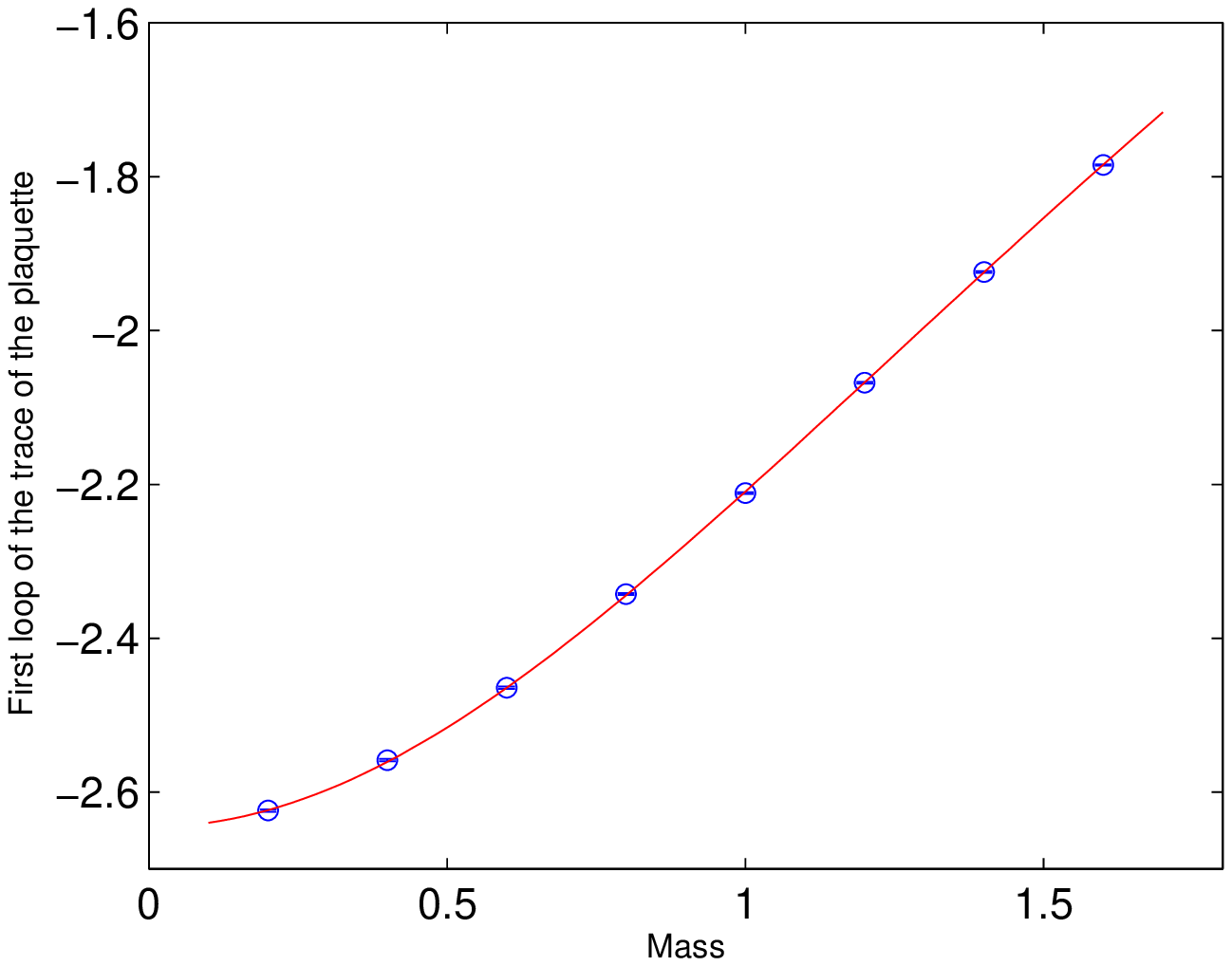, scale=0.5}
   \caption{The 1-loop trace of the plaquette vs. mass (for $L=5$): 
   the numerical results (blue dots) agree with the analytical red curve.}
   \label{Fig.1}
  \end{center}
 \end{minipage}
 \hfill
 \begin{minipage}[t]{.45\textwidth}
  \begin{center} 
   \epsfig{file=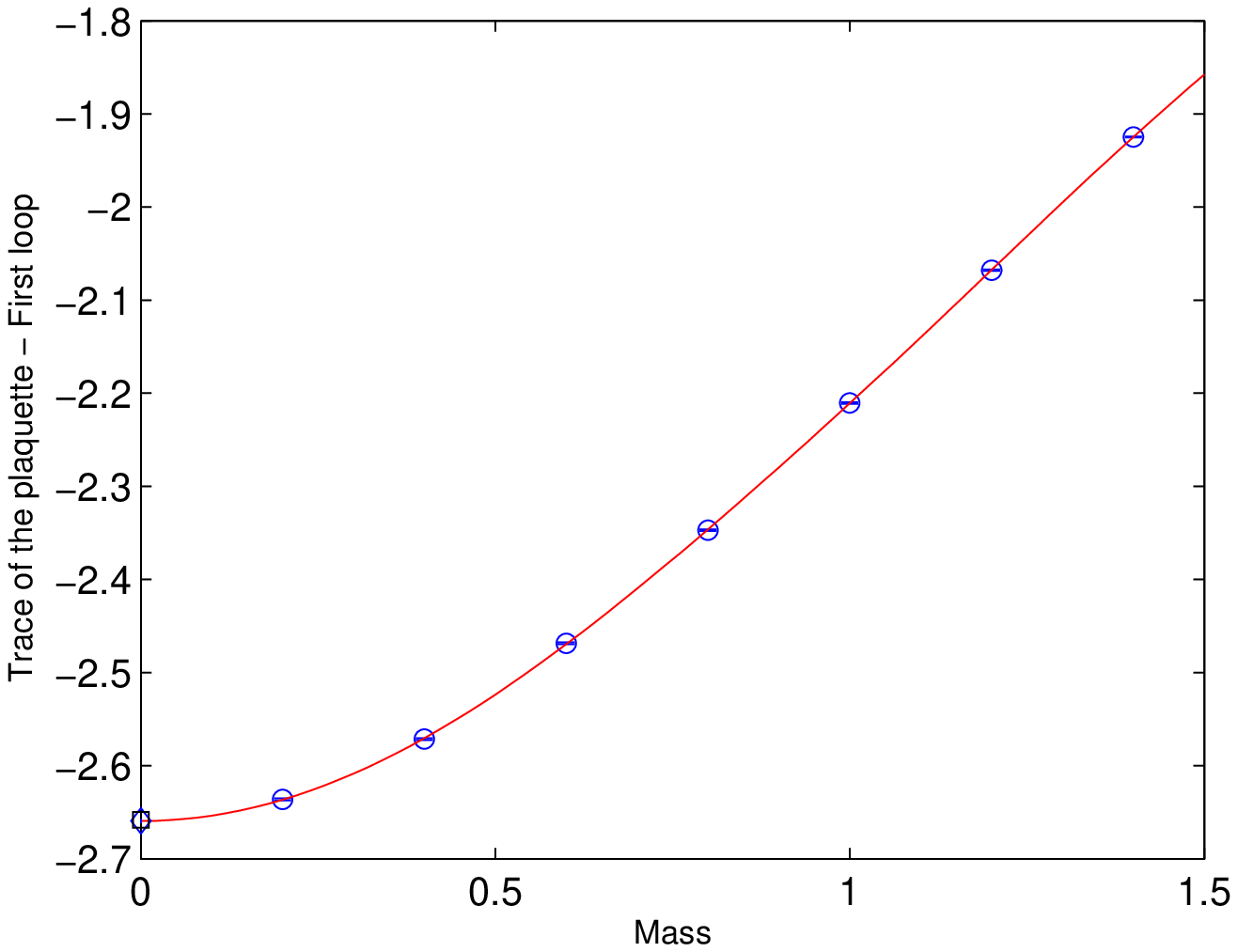, scale=0.5}
   \caption{The 1-loop trace of the plaquette vs. mass (for $L=7$): 
   the fitted curve in red approaches the expected value at $m=0$.}
   \label{Fig.2}
  \end{center}
 \end{minipage}
 \hfill
\end{figure}
%
%%%%%%%%%%%%%%%%%%%%%%%%%%%%%%%%%%%%%%%%%%%%%%%%%%%%%%%%%%%%%%%%%%%%%%%%%%%

%%%%%%%%%%%%%%%%%%%%%%%%%%%%%%%%%%%%%%%%%%%%%%%%%%%%%%%%%%%%%%%%%%%%%%%%%%%
%
\begin{figure}[h]
 \hfill
 \begin{minipage}[t]{.45\textwidth}
  \begin{center} 
   \epsfig{file=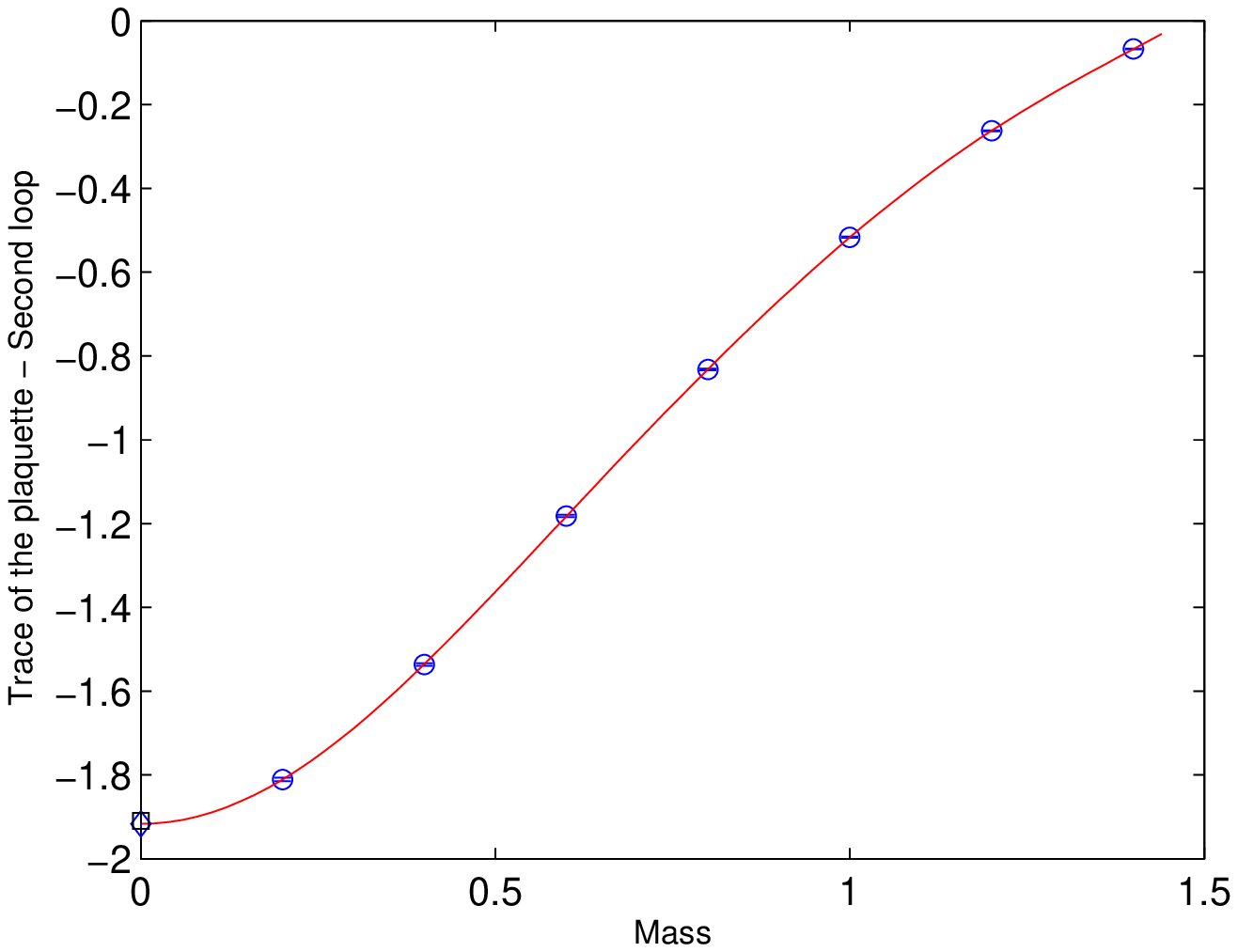, scale=0.5}
   \caption{The 2-loop trace of the plaquette (for $L=7$), 
   together with a polynomial fit in $m^2$.}
   \label{Fig.3}
  \end{center}
 \end{minipage}
 \hfill
 \begin{minipage}[t]{.45\textwidth}
  \begin{center} 
   \epsfig{file=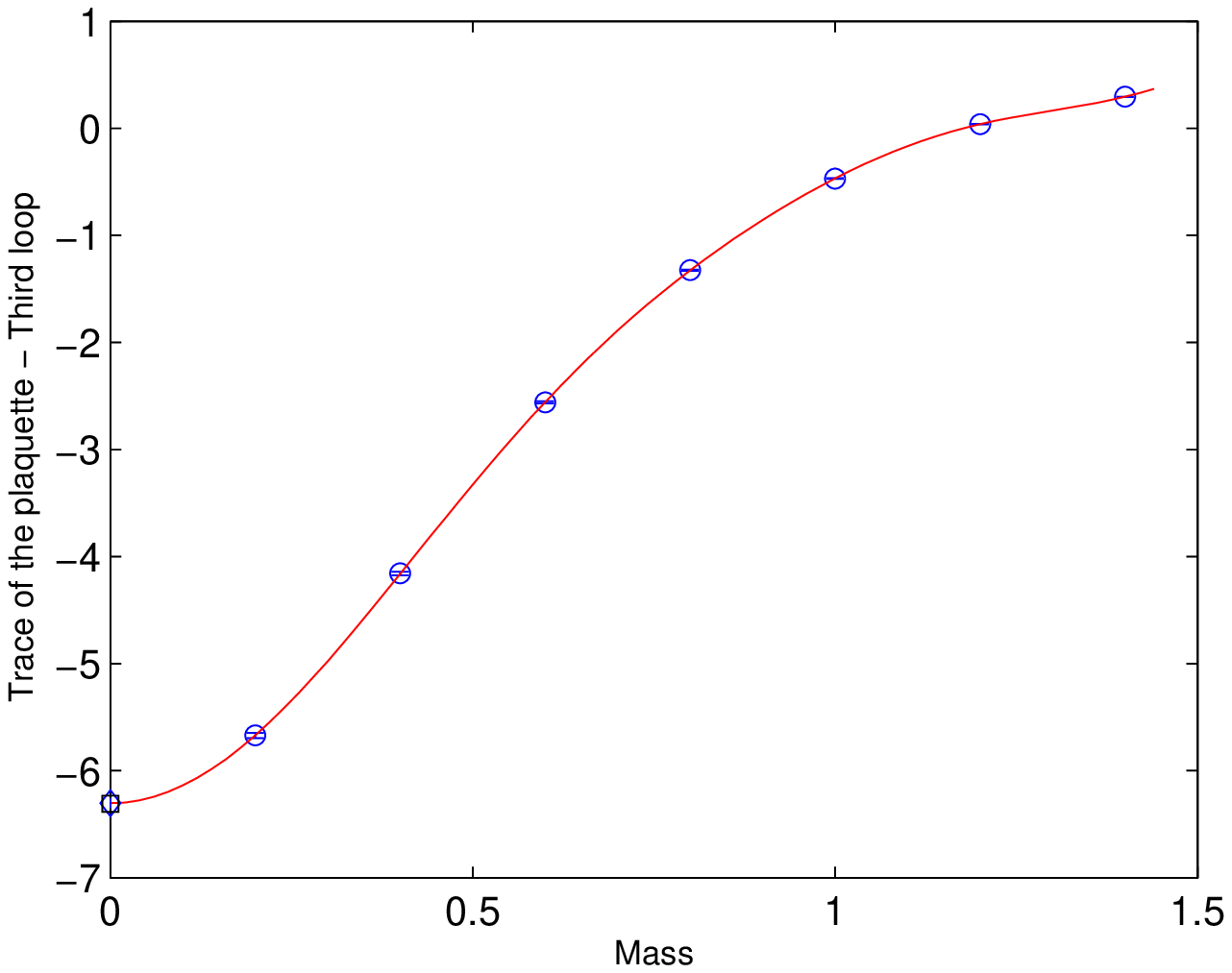, scale=0.5}
   \caption{The 3-loop trace of the plaquette (for $L=7$),
   together with a polynomial fit in $m^2$.}
   \label{Fig.4}
  \end{center}
 \end{minipage}
 \hfill
\end{figure}
%
%%%%%%%%%%%%%%%%%%%%%%%%%%%%%%%%%%%%%%%%%%%%%%%%%%%%%%%%%%%%%%%%%%%%%%%%%%%

%%%%%%%%%%%%%%%%%%%%%%%%%%%%%%%%%%%%%%%%%%%%%%%%%%%%%%%%%%%%%%%%%%%%%%%%%%
%
\begin{figure}[h]
 \hfill
 \begin{minipage}[t]{.45\textwidth}
  \begin{center} 
   \epsfig{file=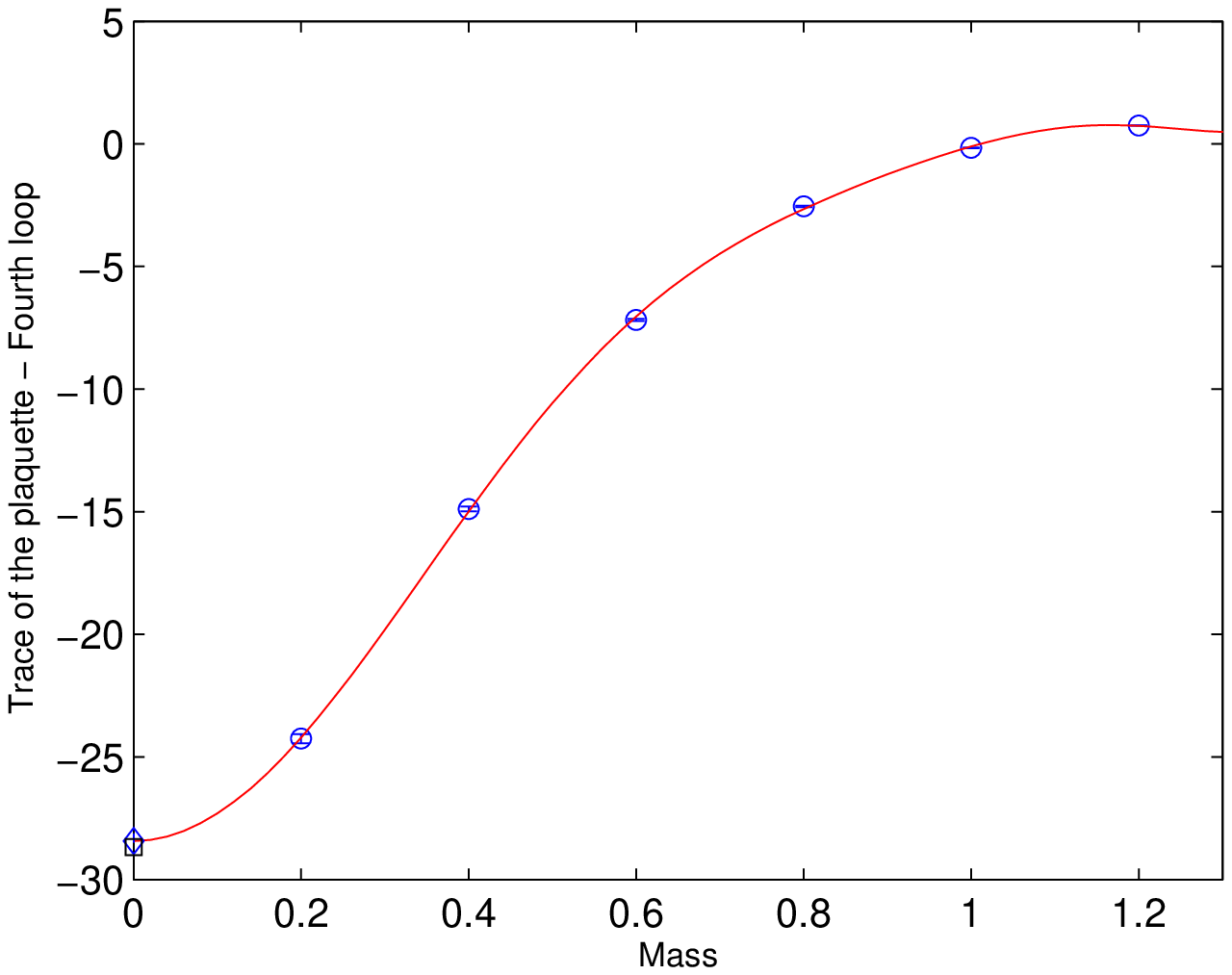, scale=0.5}
   \caption{The 4-loop trace of the plaquette (for $L=7$),
   together with a polynomial fit in $m^2$.}
   \label{Fig.5}
  \end{center}
 \end{minipage}
 \hfill
 \begin{minipage}[t]{.45\textwidth}
  \begin{center} 
   \epsfig{file=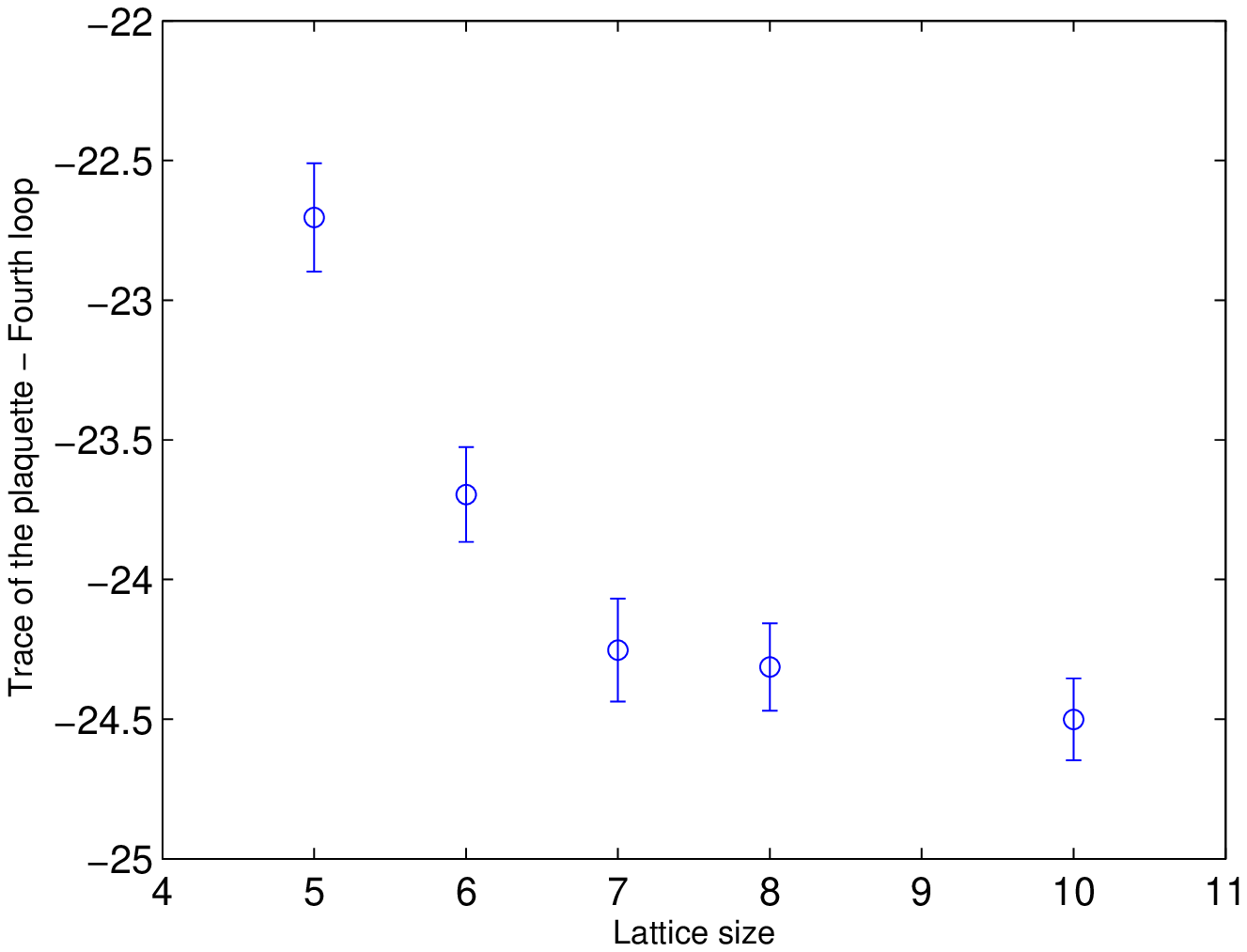, scale=0.5}
   \caption{The 4-loop trace of the plaquette vs. lattice size,
   for a fixed mass $m=0.2$.}
   \label{Fig.6}
  \end{center}
 \end{minipage}
 \hfill
\end{figure}
%
%%%%%%%%%%%%%%%%%%%%%%%%%%%%%%%%%%%%%%%%%%%%%%%%%%%%%%%%%%%%%%%%%%%%%%%%%%

%%%%%%%%%%%%%%%%%%%%%%%%%%%%%%%%%%%%%%%%%%%%%%%%%%%%%%%%%%%%%%%%%%%%%%%%%%%
%
\section{Conclusions and prospects}

It is worth stressing once again that our approach
has successfully passed the reliability checks we adopted: known
zero-mass limits are reproduced through an extrapolation, and 
the volume dependence at a fixed mass appears to disappear 
once the dimensionless combination $mL$, where 
$m$ is the mass and $L$ the lattice extent, 
is large enough. 

In order to obtain the asymptotic large-volume value at a fixed mass, 
it is still necessary to collect more statistics on bigger
lattices (for example, $L=12$ and 14) at least for the two or three
smallest masses. Then, the fitting function should be a combination 
of a negative exponential and polynomials in $mL$, 
as explained for instance in Ref.~\cite{mm}.

After subtracting the logarithmic divergence from the fitted 
infinite-volume values, the subsequent extrapolation to zero mass 
does not appear to be troublesome, given that tests with 
lower loop orders have produced good results so far.

%%%%%%%%%%%%%%%%%%%%%%%%%%%%%%%%%%%%%%%%%%%%%%%%%%%%%%%%%%%%%%%%%%%%%%%%%%%%
%

\end{document}